\documentclass[a4paper,12pt]{article}

\usepackage [T1]{fontenc}
\usepackage {calligra}
\usepackage [T1]{pbsi}
\usepackage{CJKutf8} 
\usepackage[colorlinks=true,linkcolor=blue]{hyperref}
\usepackage{amsmath,amssymb,amsfonts} 
\usepackage{graphicx}                 
\usepackage{color}                    
\usepackage{hyperref}                 
\usepackage{cite}
\definecolor{red}{rgb}{1,0,0}           
\definecolor{green}{rgb}{0,1,0}
\definecolor{blue}{rgb}{0,0,1}
\definecolor{darkblue}{rgb}{0,0,0.5}
\definecolor{lightblue}{rgb}{.5,.5,1}
\definecolor{lightgray}{gray}{.87}          
\definecolor{Dark}{gray}{.20}
\definecolor{pink}{rgb}{.95,0.82,0.92}  
\definecolor{yellow}{rgb}{1,1,0}
\definecolor{lightyellow}{rgb}{1,1,.5}
\definecolor{purple}{rgb}{0.7,0,0.85}
\definecolor{darkgreen}{rgb}{0,0.5,0}
\definecolor{orange}{rgb}{0.8,0.2,0.2}
\def \be {\begin{equation}}
\def \ee {\end{equation}}
\def \bea {\begin{eqnarray}}
\def \eea {\end{eqnarray}}

\def \rr {\raise.35ex\hbox{\small $\prime$}\kern-.17em{\mbox{\large $\imath$}}}
\def \del {\partial}
\def \dels {\partial\kern-.5em / \kern.5em}
\def \As {{A\kern-.5em / \kern.5em}}
\def \Ds {D\kern-.7em / \kern.5em}

\def \eps {\epsilon}

\newcommand{\detail}[1]{}

\newcommand{\hide}[1]{}

\begin{document}

\pagestyle{plain}

\begin{CJK}{UTF8}{bsmi}

\begin{titlepage}

\begin{center}

\noindent
\textbf{\LARGE
Comment on Self-Consistent Model of 
\vskip.6cm
Black Hole Formation and Evaporation
}
\vskip .5in
{\large 
Pei-Ming Ho\footnote{e-mail address: pmho@phys.ntu.edu.tw}
}
\\
{\vskip 10mm \sl
Department of Physics and Center for Theoretical Sciences, \\
Center for Advanced Study in Theoretical Sciences, \\
National Taiwan University, Taipei 106, Taiwan,
R.O.C. 
}\\

\vskip 3mm
\vspace{60pt}
\begin{abstract}

In an earlier work,
Kawai et al proposed a model of black-hole
formation and evaporation,
in which the geometry of 
a collapsing shell of null dust is studied,
including consistently the back reaction of its Hawking radiation.
In this note,
we illuminate the implications of their work, 
focusing on the resolution of the information loss paradox
and the problem of the firewall.

\end{abstract}
\end{center}
\end{titlepage}

\setcounter{page}{1}
\setcounter{footnote}{0}
\setcounter{section}{0}


\section{Introduction}

The information loss paradox \cite{Hawking:1976ra}
has intrigued theoretical physicists
since Hawking's discovery that black holes evaporate.
The unitarity of quantum mechanics 
demands that Hawking radiation be not thermal,
and that it carry information of the matter inside the horizon.
However, 
to propose a mechanism to transfer information 
from the collapsed matter to Hawking radiation,
there are various theoretical obstacles 
such as the no-hair theorem, 
the no-cloning theorem, the monogamy of entanglement,
as well as the causality and locality of semiclassical effective theories.
An important progress was the proof of the small-correction theorem \cite{Mathur:2009hf},
which states that the information loss cannot be recovered by 
anything less than order-one correction at the horizon.
See Ref.\cite{Mathur:2009hf} 
for a more precise explanation of the information loss paradox.

As Hawking radiation is originated from the horizon,
the collapsing matter has to leave all of its information 
before entering the horizon
(unless the information can travel non-locally
from the inside to the outside of the horizon at a later time).
On the other hand, 
it is commonly believed that
nothing special should happen to an infalling observer at the horizon,
and so the information should be carried into the horizon.
The black-hole complementarity \cite{Susskind:1993if,'tHooft:1984re} 
is thus proposed as a resolution of the puzzle.

\begin{figure}
\vskip-4em
\includegraphics[scale=0.5]{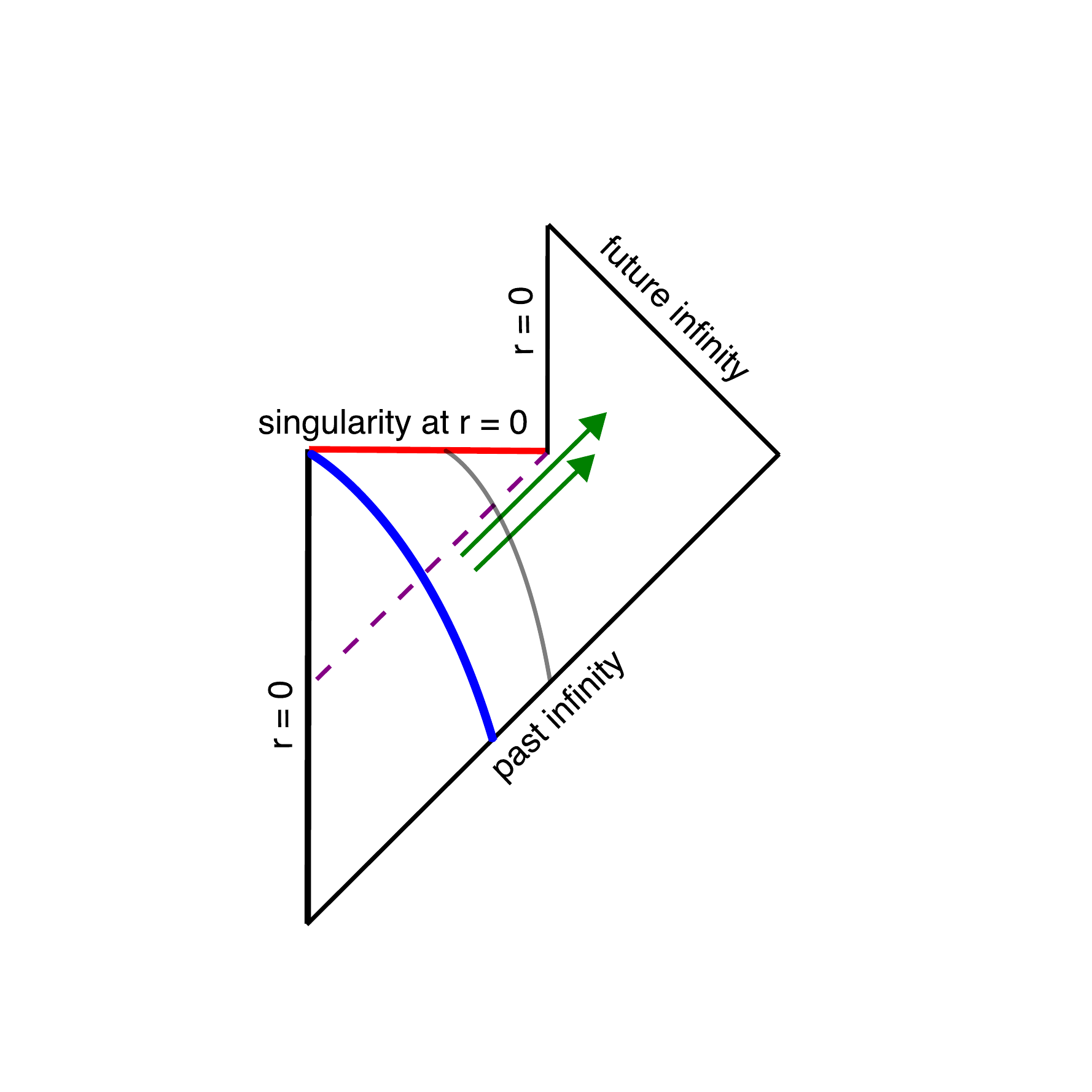}
\vskip-3em
\label{Penrose-old-1}
\caption{\small
Penrose diagram for the conventional model 
of the formation and evaporation of a black hole:
The blue curve represents the collapsing matter,
the green arrows Hawking radiation,
and the red line the singularity at the origin.
The horizon is represented by the dash line,
and an infalling observer the gray curve.
}
\vskip1em
\end{figure}

The puzzle can be presented through
the Penrose diagram for the conventional model 
of the formation and evaporation of a black hole.
(See Fig. \ref{Penrose-old-1}.)
An object falling in through the horizon ends up 
at the singularity at $r = 0$,
while all of its information must come out 
of the horizon through Hawking radiation 
to a distant observer to avoid information loss.
The puzzle about information arises 
essentially because there is a region 
(the region behind the horizon) 
from which outgoing light cones
can not reach to the future infinity.

On the other hand,
a different model of the (non-)formation and evaporation
of a black hole was presented in the work of Kawai, Matsuo and Yokokura \cite{Kawai:2013mda},
which can be directly taken as a resolution of the information loss paradox.
We will refer to their model \cite{Kawai:2013mda} as the KMY model.
It is the purpose of this article to 
illuminate the implications of their work 
on problems about the information loss paradox.

The key message of the KMY model is that
the back reaction of Hawking radiation
must be taken into consideration before 
the black hole forms.
\footnote{
The Hawking radiation occurring before the formation of a black hole 
is sometimes called a pre-Hawking radiation.
}
The formation and evaporation of a black hole 
happen at the same time as a single process.
Hawking radiation appears before there is a horizon,
and as it takes away energy from the system,
its back reaction makes it harder for the horizon to emerge.
In fact, 
it was shown \cite{Kawai:2013mda} that
no horizon ever appears
(unless it is already there in the initial state),
not only from the viewpoint of a distant observer,
but also from the viewpoint of an infalling observer.
The collapsing matter never completely falls inside the Schwarzschild radius.
As a result, 
the conventional assumption of the empty horizon of a black hole
is replaced by the presence of collapsing matter
around the Schwarzschild radius,
and the information loss paradox is resolved.
Similar viewpoints were taken in \cite{Callan:1992rs,Barcelo:2007yk},
although the reasoning and arguments are different.

Since there will never be a horizon,
strictly speaking there is no black hole.
In the literature,
a to-be-formed black hole is called
an incipient black hole,
which will sometimes also be simply called a black hole
in this note.
\footnote{
An incipient black hole can
also have a dark surface and Hawking radiation, 
just like a real black hole.
}

In this note,
we will point out the implication of the KMY model that
there is no horizon regardless of the nature of the collapse of matter.
We will give a proof of no horizon
based on semiclassical calculations.
We will also point out another implication of the KMY model,
which reduces the firewall \cite{Almheiri:2012rt}
to a weak radiation.

Incidentally,
one may wonder what happens to a static black hole.
A static black hole can be constructed  
through an adiabatic process in a heat bath,
with the heat inflow balanced by Hawking radiation,
and it was found that there is still no horizon
\cite{Kawai:2013mda,Kawai:2014afa}.
We will not discuss this case in this note.

\section{Geometry Outside Collapsing Shell}
\label{outside}

It is well known that,
from the viewpoint of an observer outside the black hole,
the motion of an infalling particle gets slower
as it gets closer to the horizon,
and it can never cross the horizon.
Due to Hawking radiation, 
the black hole evaporates and disappears 
within a finite amount of time.
Therefore, 
for a distant observer,
the black hole evaporates before 
the infalling particle reaches the horizon.
This does not however imply that 
an infalling object cannot cross the horizon 
within finite proper time.
According to the Penrose diagram 
of the traditional view of the black-hole space-time
(Fig. \ref{Penrose-old-1}),
an infalling object can pass through the horizon in finite proper time.
But to a distant observer 
it never reaches the horizon until 
the horizon shrinks to zero so that
it touches the singularity at the origin 
at the same moment when the black hole 
is completely evaporated.

The KMY model claims,
however,
that if the effect of the back reaction of 
Hawking radiation is included from the very beginning,
the Penrose diagram has to be significantly modified.
An infalling observer can never pass the horizon,
as there is no horizon.

The KMY model describes
the collapse of a spherical matter shell,
and the corresponding space-time geometry.
Both the collapsing shell and the Hawking radiation 
are approximated by spherical configurations.
It also ignores massive particles in the Hawking radiation.
But it includes the back-reaction of Hawking radiation on the geometry,
and thus the time-dependence of the Schwarzschild radius.

In this section,
we study the space-time geometry outside the shell 
in the semiclassical approximation.
The energy-momentum tensor in the Einstein equation
\be
G_{\mu\nu} = 8\pi \langle T_{\mu\nu}\rangle
\label{Einstein}
\ee
is given by the expectation value of the quantum energy-momentum operator
of matter fields,
which is identified with the Hawking radiation outside the collapsing shell.
It is assumed that there is nothing but Hawking radiation
outside the spherical matter shell.

Under these assumptions,
the most general solution to Einstein's equation 
for the outside of a collapsing spherical shell 
is the outgoing Vaidya metric \cite{Vaidya:1951zz}:
\footnote{
Both the ingoing and outgoing Vaidya metrics have been 
applied to the problem of black-hole formation and evaporation 
in the literature.
See, e.g.
\cite{Hiscock,Kuroda}.
}
\be
ds^2 = - \left(1-\frac{a(u)}{r}\right)du^2 - 2du dr
+ r^2 d\Omega^2.
\label{outgoing-Vaidya}
\ee
In the units $G = c = 1$,
the Bondi mass for this metric is $M(u) = a(u)/2$
and the only non-vanishing component of the Einstein tensor is
\be
G_{uu} = - \frac{\dot{a}(u)}{r^2}.
\ee
This corresponds to an energy-momentum tensor
with the only non-vanishing component
\be
T_{uu} = \frac{1}{8\pi} G_{uu} = - \frac{1}{8\pi}\frac{\dot{a}(u)}{r^2}.
\label{Tuu}
\ee
The weak energy condition demands that $\dot{a} < 0$.

The outgoing Vaidya metric has been known as 
a solution to Einstein's equation
for a spherical star emitting null dust.
Referring to the outer radius of the collapsing shell as $R(u)$,
the metric (\ref{outgoing-Vaidya}) should be valid 
for the outside of a spherical collapsing shell for all $r \geq R(u)$,
assuming that there are only massless particles in Hawking radiation
\cite{Kawai:2013mda}.

Recall that the geometry outside a static mass at the origin is given by
the Schwarzschild solution:
\be
ds^2 = - \left(1-\frac{a_0}{r}\right) dt^2 
+ \left(1-\frac{a_0}{r}\right)^{-1} dr^2
+ r^2 d\Omega^2,
\label{BH-rt}
\ee
where the Schwarzschild radius is
\be
a_0 = 2M,
\ee
and $M$ is the total mass.
It can also be rewritten 
in terms of the outgoing Eddington-Finkelstein coordinates as
\be
ds^2 = - \left(1-\frac{a_0}{r}\right)du^2 - 2du dr
+ r^2 d\Omega^2,
\ee
where 
\be
u = t - r^*(r),
\qquad
r^*(r) \equiv r + a_0 \log\left|\frac{r}{a_0} - 1\right|,
\label{r*}
\ee
through a change of coordinates.
This metric is in the form of the outgoing Vaidya metric (\ref{outgoing-Vaidya}),
and we will refer to $a(u)$ in (\ref{outgoing-Vaidya}) as the Schwarzschild radius.
As energy is radiated away through Hawking radiation,
the Bondi mass $M(u)$ decreases with time,
and so is the Schwarzschild radius $a(u)$.

Let us emphasize here that, 
despite the connection between the outgoing Vaidya solution
and the Schwarzschild solution,
whether there is a horizon for the outgoing Vaidya metric (\ref{outgoing-Vaidya})
is still a question to be answered.
The initial state of the collapsing shell is assumed to 
satisfy $R(u_1) > a(u_1)$ at the initial time $u = u_1$.
Since the shell continues collapsing,
and the energy flux in Hawking radiation is tiny for a large initial mass $M(u_1)$,
it is natural to expect that at some later point
a horizon would appear when 
the shell shrinks to a size smaller than the Schwarzschild radius.
Our task is to examine carefully 
whether this naive expectation is really what 
Einstein's equation tells us through the outgoing Vaidya metric.

If there really would be a horizon, 
at least one of the two following things must happen:
(i) $R(u) < a(u)$ at a later time $u > u_1$; 
\footnote{
We do not have to assume that $r = a(u)$ is the horizon.
But if the horizon exists,
$R(u)$ will go to zero after passing through the horizon,
so that $R(u)$ should not have a lower bound 
at any finite value such as $a(u)$.
}
or
(ii) the outgoing Vaidya metric (\ref{outgoing-Vaidya}) 
is geodesically incomplete, 
so that a horizon can hide behind the infinity $u = \infty$.
We will show in the following that
neither of the two things happen, 
so there is no horizon,
contrary to the naive expectation.

\section{Evaporation}

One may wonder if Hawking radiation would still appear
if there is no horizon.
We argue that,
as the collapsing shell gets very close to the Schwarzschild radius,
the geometry just outside the shell is indistinguishable 
from the near horizon region of a real black hole
and thus Hawking radiation is expected to appear,
although its spectrum can be modified.
This view is supported by the literature \cite{Hawking-Radiation}.

In fact, 
it has been calculated in this context in \cite{Kawai:2013mda},
in the absence of a horizon.
Hawking radiation appears on the shell
as well as outside the shell.
The energy flux in Hawking radiation determines 
the rate of change in the mass $\dot{M}(u) = \dot{a}(u)/2$.
In 4 dimensions,
the Hawking temperature $T_H(u)$ is roughly $1/a(u)$,
\footnote{
This is an approximation of a more precise expression \cite{Kawai:2013mda}
for large $a$.
}
and the energy flux in the radiation of massless particles is
\cite{Kawai:2013mda}
\be
J \propto \mbox{Area}\times T_H^4 \propto \frac{1}{a^2(u)}.
\label{Ju1}
\ee
Thus in the semiclassical, spherical symmetry approximation,
\be
\dot{a}(u) = - C/a^2(u)
\label{dota}
\ee
for some constant $C$.
($C \equiv \frac{N}{48\pi}$,
which is proportional to the number $N$ of species of massless particles.)
The solution is 
\be
a(u) = 
\left\{
\begin{array}{ll}
3C^{1/3}(u_0-u)^{1/3} & (u < u_0) \\
0 & (u \geq u_0)
\end{array}
\right.
\label{au}
\ee
for some constant $u_0$
when the matter shell is completely evaporated.
The only information we need from this calculation 
is that the Schwarzschild radius $a(u)$
decreases monotonically to zero in a finite elapse in $u$,
and after that the spacetime becomes Minkowskian.
The argument below for the absence of horizon 
will not rely on the detail of this solution (\ref{au}).

\hide{
The expression of the energy flux $J$ used above
is an approximation of the more accurate expression \cite{Kawai:2013mda}
\be
J = \frac{N}{96\pi}\left(\frac{1}{a^2(u)}+ \frac{4}{a^2(u)}\frac{da}{du}\right),
\label{Ju2}
\ee
which leads to a deviation of the solution (\ref{au}).
In the limit $u \rightarrow 0$,
\be
a(u) \simeq \frac{u_0-u}{4},
\label{a-smallu}
\ee
independent of $N$.
After the mass $M(u) = a(u)/2$ goes to $0$ at $u_0$,
the matter shell is completed evaporated.
}

Of course,
the semiclassical analysis of Hawking radiation
is not expected to be valid all the way to the Planck scale.
In this work 
we are not really concerned with the fate of the collapsing shell 
when it is reduced to a Planck size.
Our aim is to understand the semiclassical physics involved
to explain the formation and evaporation of 
an astronomical black hole.
We will say that a macroscopic black hole 
is completely evaporated even when there are 
Planck-scale remnants
(see e.g. \cite{YC}),
whose entropy is negligible compared with
an astronomical massive object.
In this sense 
we assume that an incipient black hole evaporates completely
within finite $u$.

Notice that it is generally assumed that
a macroscopic black hole evaporates away in finite time $t$.
At large $r$ ($r \gg a(u)$),
we have $dt \simeq du + dr$ for the proper time $t$
of a distant observer located at a fixed $r$,
hence an elapse in finite $t$ means the same as finite $u$.

\section{Proof of No Horizon}

To prove that there is no horizon,
we will show the following two things:
(i) $R(u) > a(u)$ for all $u > u_1$ 
as long as $R(u_1) > a(u_1)$ at an initial time $u = u_1$,
until the collapsing shell is completely evaporated;
and
(ii) when the collapsing shell is completely evaporated,
$R(u) > 0$ and extends into the Minkowski space,
which is geodesically complete,
so that a horizon cannot hide behind the infinity $u = \infty$.

First we sketch a proof by contradiction 
to prove that the outer radius of the collapsing shell 
is always outside the Schwarzschild radius,
in the context of semiclassical theories 
with minimal assumptions.

Assume that $R(u)$ eventually falls inside $a(u)$.
The outer radius would have to pass through the Schwarzschild radius,
with the velocity of the outer surface ($r=R(u)$)
larger (moving faster towards the origin)
than the velocity of the Schwarzschild radius ($r=a(u)$)
at a certain instant $u = u'$ when $R(u') = a(u')$.
The outgoing Vaidya metric (\ref{outgoing-Vaidya})
would be applicable to the point $r = a(u)$ for $u \geq u'$,
and the trajectory of the Schwarzschild radius
($r = a(u)$) would be space-like:
\footnote{
Let us emphasize that we are showing 
a proof by contradiction.
We are not saying that 
the trajectory $r = a(u)$ is actually space-like,
as the outgoing Vaidya metric is not applicable to it
with $R(u) > a(u)$.
In the next section,
we will show more directly that 
$R(u) > a(u)$ at all times.
}
\be
ds^2 = - 2 du da(u) = - 2 du^2 \dot{a}(u) > 0
\qquad (u \geq u')
\ee
because $\dot{a} < 0$.
As the trajectory of a point on the outer surface 
of the shell ($r = R(u)$) is either light-like or time-like,
it is impossible for the outer surface of the shell
to move faster towards the origin than the Schwarzschild radius,
and thus we have a contradiction.

In other words, 
if $R(u) > a(u)$ for an initial state at $u = u_1$,
it is impossible to have $R(u) < a(u)$ at a later time $u > u_1$.
The radius $R(u)$ of the collapsing shell 
can never catch up with the Schwarzschild radius $a(u)$,
and the horizon can never emerge,
at least when the outgoing Vaidya metric (\ref{outgoing-Vaidya})
is still valid.

\begin{figure}
\vskip-0em
\includegraphics[scale=0.5]{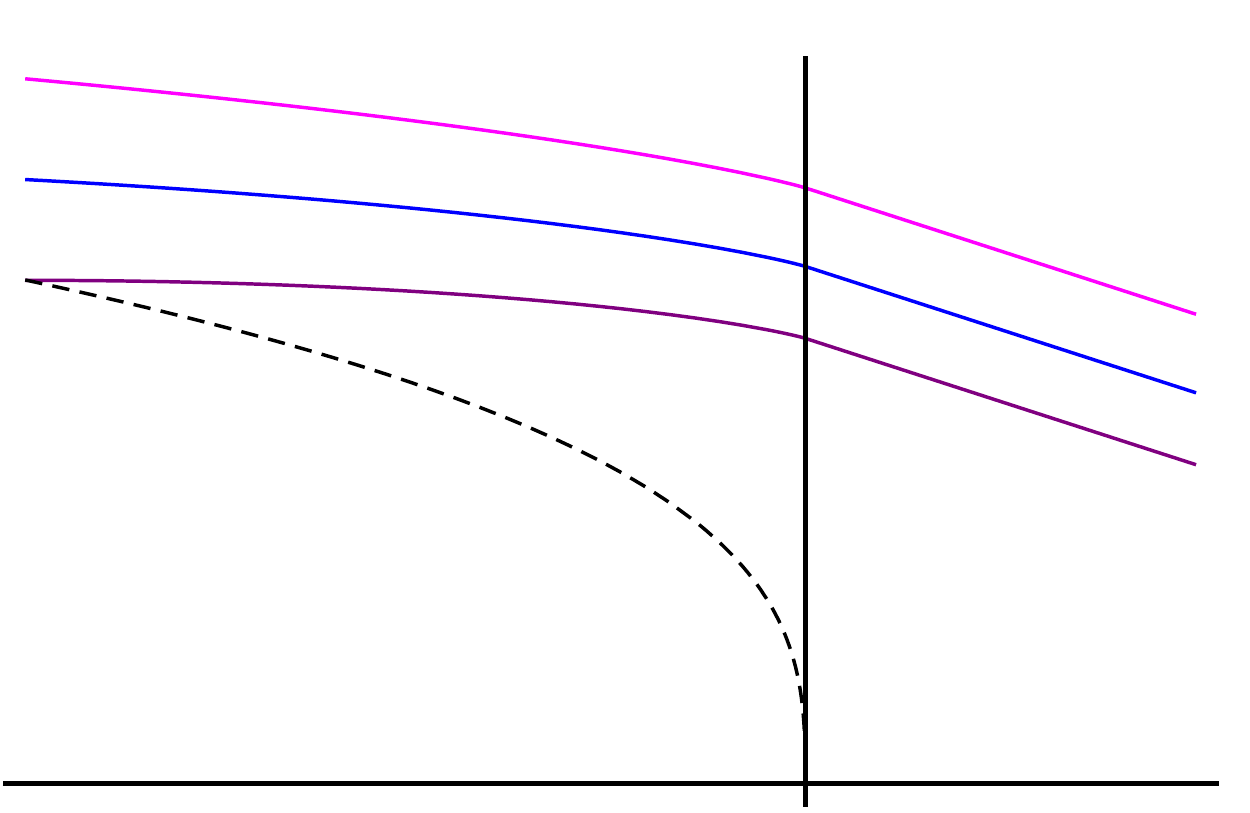}
\hskip5em
\includegraphics[scale=0.5]{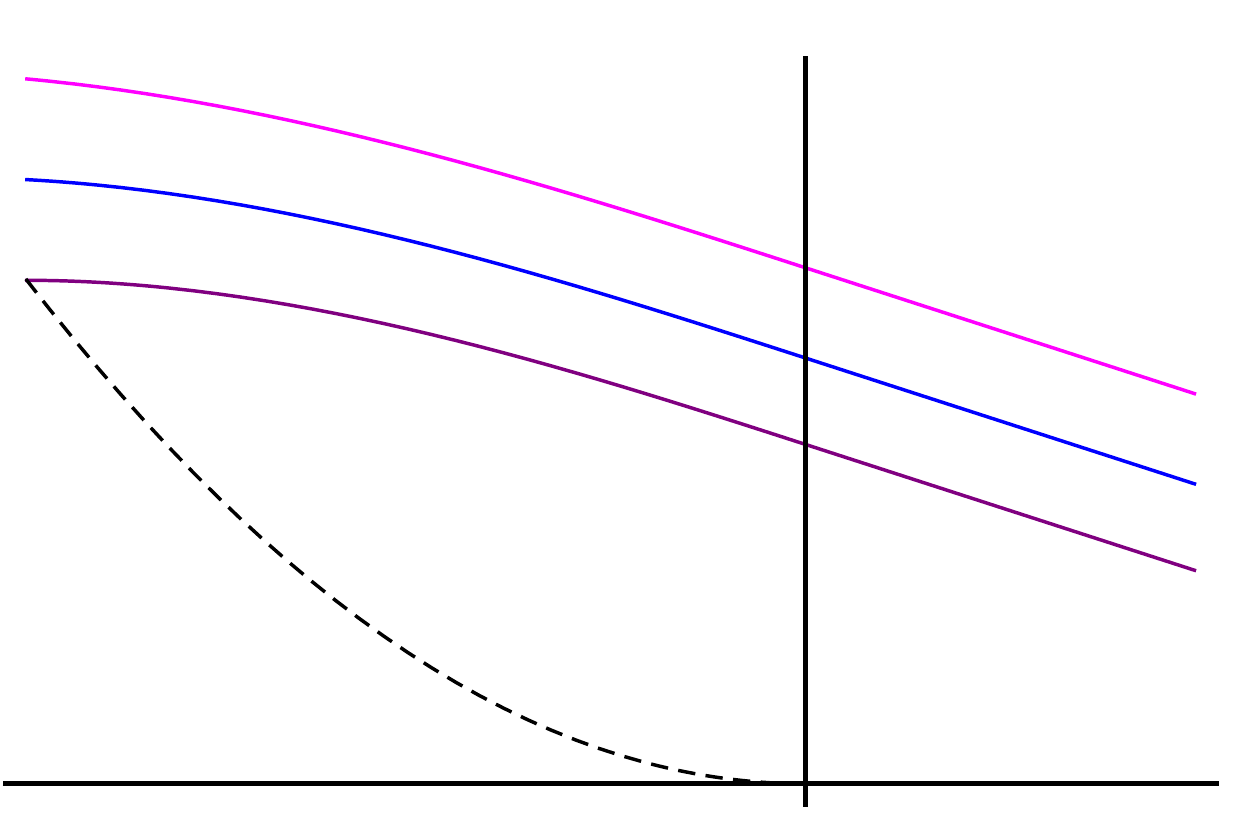}
\vskip0em
\hskip3.5cm (a) \hskip8cm (b)
\caption{\small In both figures,
the horizontal axis is $u$,
and the vertical axis for $r$ is located at $u=u_0$,
where the matter shell is completely evaporated,
and the spacetime is Minkowskian for $u \geq u_0$.
The dashed curve represents $a(u)$,
and the solid curves $R(u)$ for different initial conditions.
The two figures differ in the trajectory of $a(u)$.
It is $a(u) \propto (u_0-u)^{1/3}$ in (a) 
and $a(u) \propto (u_0-u)^2$ in (b).
It is a robust feature that 
$R(u)$ remains at a finite value at $u_0$
when $a(u)$ goes to zero.}
\label{R-trajectories}
\vskip1em
\end{figure}

The outgoing Vaidya metric remains valid as
it turns into the Minkowski space metric when $a(u) = 0$
after the shell disappears at $u = u_0$.
In Fig. \ref{R-trajectories},
we demonstrate the trajectories of $R(u)$ for different initial conditions,
assuming that the outer radius of the matter shell 
shrinks at the highest speed possible --
the speed of light,
which is still slower than $a(u)$.
Hence,
for arbitrary initial conditions,
$R(u_0) > 0$ and it extends into the Minkowski space smoothly.
As the Minkowski space has no horizon and is geodesically complete,
the trajectory of $R(u)$ is geodesically complete 
without crossing the Schwarzschild radius anywhere.
At a speed lower than or equal to light, 
an infalling observer originally outside the collapsing shell
would therefore also have a geodesically complete trajectory
without crossing a horizon
(or hitting any singularity).
We conclude that 
there can be no horizon at all 
for a collapsing matter shell no matter how fast it collapses.

A few assumptions were made in the argument above.
The use of the Vaidya metric has to do with 
the fact that the emission of massive particles in Hawking radiation is ignored.
But it is a robust feature that
the trajectory $r = a$ of the Schwarzschild radius is space-like
when $\dot{a} < 0$.
This is because the trajectory $r = a_0$ is null-like 
when $a = a_0$ is constant.
If we had used the metric (\ref{BH-rt})
and replaced $a_0$ by $a(t)$ with $\dot{a}(t) < 0$,
we could still show that,
for any finite $|\dot{a}(t)| \neq 0$,
the trajectory $r = a(t) + \epsilon$ 
of a point just outside the Schwarzschild radius
is space-like
for sufficiently small $\epsilon > 0$.
Furthermore,
the inclusion of massive particles in Hawking radiation
would lead to a faster evaporation,
making it even harder for $R(u)$ to catch up with $a(u)$.

In addition to spherical symmetry,
we have also assumed that $\dot{a} < 0$
due to Hawking radiation until $a(u_0) = 0$,
and that the shell evaporates completely at a finite value of $u = u_0$.
The value of $|\dot{a}|$ can otherwise be arbitrarily small.


\hide{
The argument partially relies on the use
of a metric of the form of the outgoing Vaidya metric
(see \cite{Wang:1998qx} for the generalized Viadya metric),
which can be easily generalized.
When the radiation of massive particles is taken into account,
the Vaidya metric has to be modified,
but we believe that the basic ingredients of the argument will persist.
}

Note that we have not made any assumption about the constituents of the matter shell,
except that its trajectory is time-like or light-like,
unlike other works with a similar conclusion \cite{Mersini-Houghton:2014zka}.

\section{Geometry of Collapsing Shell}

Let us give more details for the simple example of
a collapsing matter shell composed of massless dust \cite{Kawai:2013mda}.
The special case of null dust is interesting because
if even a shell collapsing at the speed of light cannot form a horizon,
a generic time-like collapsing shell most certainly cannot, either.

For spherically symmetric configurations,
the light-cone directions for the outgoing Vaidya metric (\ref{outgoing-Vaidya}) are given by
\be
du = 0, \qquad
\left(1-\frac{a(u)}{r}\right)du + 2 dr = 0.
\ee
The solutions to the first equation $du = 0$
give trajectories of the outgoing massless particles in Hawking radiation.
The solutions to the second equation
then describe the trajectories of infalling massless particles,
including those at the outer surface of the shell of null dust.
Therefore,
the outer radius $R(u)$ of the collapsing spherical shell
satisfies \cite{Kawai:2013mda}
\be
\frac{dR(u)}{du} = - \frac{1}{2} \left(1-\frac{a(u)}{R(u)}\right).
\label{dRdu}
\ee
According to this equation,
$dR/du \rightarrow 0$ as $R \rightarrow a$,
and so $R$ can never get inside the Schwarzschild radius.

\hide{
\begin{figure}
\vskip-0em
\includegraphics[scale=0.4]{BlackHole.pdf}
\includegraphics[scale=0.4]{CollapsingShell.pdf}
\vskip-2em
\hskip3.5cm (a) \hskip6.5cm (b)
\caption{\small 
(a) In the conventional picture of a black hole,
the infalling matter falls into the Schwarzschild radius $a$
(the dash line),
which changes slowly with time. 
(b) Including the back-reaction of Hawking radiation,
the infalling matter can never fall into the Schwarzschild radius $a(u)$.}
\label{ru-diagrams}
\vskip1em
\end{figure}
}

\begin{figure}
\vskip-0em
\includegraphics[scale=0.5]{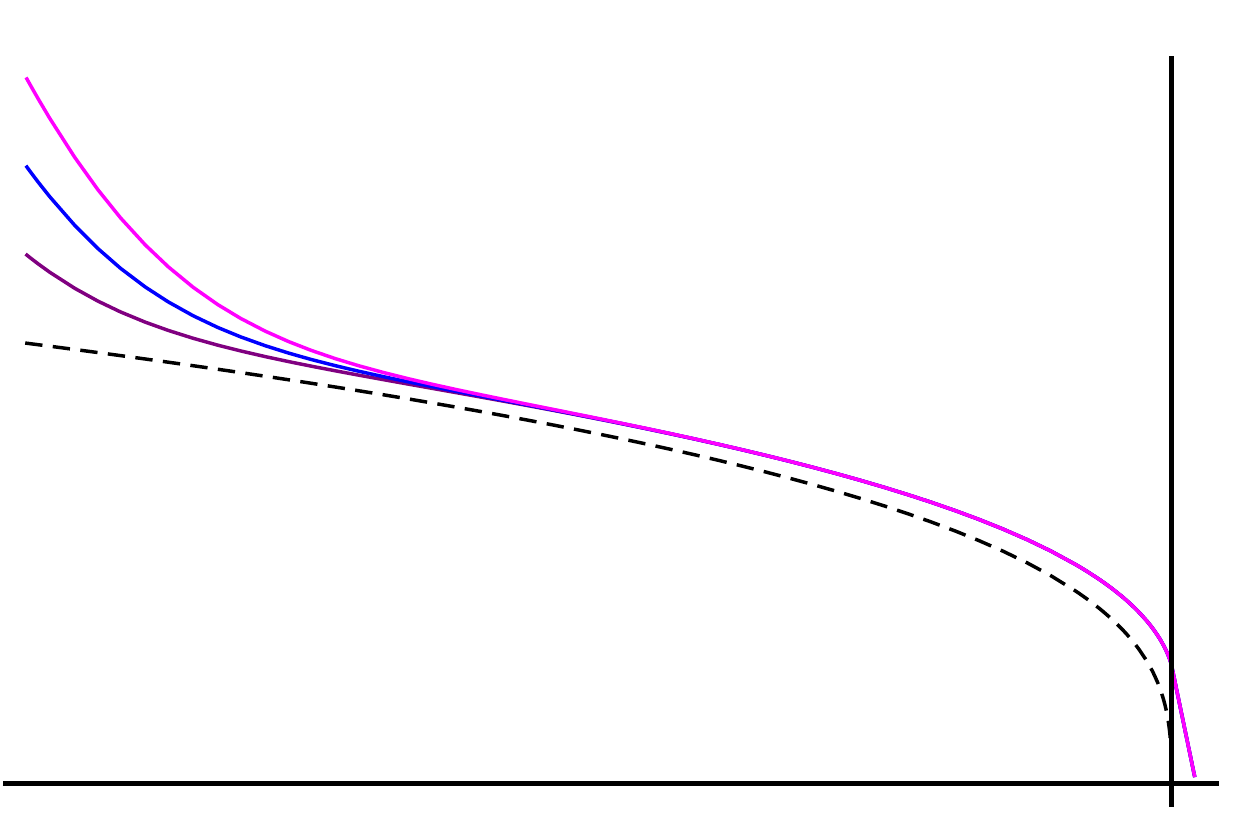}
\caption{\small 
The horizontal axis is $u$ and
the vertical axis is located at $u = u_0$,
where $a(u_0) = 0$.
The trajectories of $R(u)$ as solutions of (\ref{dRdu})
for different initial conditions are given, 
in comparison with that of $a(u)$ given by (\ref{au}) as a dashed curve.
For each trajectory of $R(u)$,
it starts from its initial value on the left
and approaches $a(u)$ quickly.
The middle part is described by (\ref{R-sol})
when $R(u)$ is very close to $a(u)$.
When $a(u)$ goes to 0,
$R(u)$ is approximated by (\ref{R-2}),
and this regime was the focus of Fig. \ref{R-trajectories}(a),
although with different initial conditions.}
\label{R-trajectory-2}
\vskip1em
\end{figure}

Samples of the null-like trajectories of $R(u)$ 
for different initial conditions at $u_1 < u_0$
are illustrated in Fig. \ref{R-trajectories}
and Fig. \ref{R-trajectory-2}.
The fact that $R(u)$ remains outside the Schwarzschild radius $a(u)$
is a robust feature that
does not rely on the explicit dependence of $a(u)$ on $u$,
but only the fact that $\dot{a} < 0$.
In general,
since the trajectory of $r = R(u)$ is null-like 
but that of $r = a(u)$ is space-like.
$R(u)$ can never catch up with $a(u)$ before $u_0$,
regardless of the initial condition.
The trajectory of $R(u)$ can always be 
smoothly extended into the Minkowski space, 
and thus the trajectory of $R(u)$ is geodesically complete.

For a sufficiently large collapsing shell,
typically $R(u)$ gets very close to the Schwarzschild radius
for a long period of time
(see Fig. \ref{R-trajectory-2}).
During the stage when
$R(u)$ is very close to $a(u)$,
\footnote{
Notice that we are studying the separation between $R(u)$
and $a(u)$ as values of the coordinate $r$,
which is not the physical distance.
This is nevertheless sufficient for our purpose
to show the absence of horizon.
}
to the first order approximation,
the solution of $R(u)$ to the equation above is \cite{Kawai:2013mda}
\be
R(u) = a(u) - 2a(u)\dot{a}(u) + \cdots.
\label{R-sol}
\ee
This is a good approximation if $|\dot{a}|\ll 1$,
when the outer radius of the shell gets close to the Schwarzschild radius
($|R-a| \ll a$),
before the matter shell evaporates towards the Planck scale
when $a \rightarrow 0$.

At the final stage of evaporation ($u \sim u_0$ but $u < u_0$),
with the Schwarzschild radius $a(u) \sim 0$,
the trajectory of $R(u)$ can be solved from
the light-cone condition (\ref{dRdu})
in the limit $u \rightarrow u_0^-$ 
(when $a(u) \ll R(u)$) by
\be
R(u) \rightarrow \frac{u_0-u}{2} + c
\label{R-2}
\ee
for some constant $c > 0$ determined by the initial condition.
(See Fig. \ref{R-trajectories}.)
Hence the matter shell remains a finite size 
when it completely evaporates.

\hide{
In comparison,
for constant $a(u) = a_0$,
the radius $R(u)$ of the collapsing shell can 
easily pass through the Schwarzschild radius
so that a horizon can emerge.
}

Since the trajectory of $R(u)$ is null-like and is geodesically complete,
it is impossible for a time-like or null-like geodesic
originated from the region outside the shell 
to pass through the Schwarzschild radius,
as we commented in the previous section.

\hide{
Let the mass enclosed inside a layer of radius $r$ 
at a fixed time $u_1$ be denoted $m(r)$.
An outside observer should see the radius of the layer
approaching to the Schwarzschild radius $2 m(r)$ at a sufficiently late time.
Assuming a sufficiently compact matter shell
with high initial density,
there may be a stage when the radii of all layers
are approximately equal to $2 m(r)$
(when the radius $R(u)$ of each layer is approximated by (\ref{R-sol})
but not (\ref{R-2})),
the distribution of mass of the collapsing shell 
can be approximated by the state 
with the mass density
\be
\rho(r) = \frac{1}{8\pi r^2},
\label{rho}
\ee
so that
\be
m(r) = \int_0^r dr' \; 4\pi {r'}^2 \rho(r') = r/2.
\label{mr}
\ee
From the viewpoint of a distant observer,
the collapsing shell evolves towards a state 
with the universal density $\rho(r)$ (\ref{rho})
for $r < R(u)$.

For the region where the mass distribution 
is approximated by (\ref{mr}),
the metric (\ref{outgoing-Vaidya}) is 
\be
ds^2 \simeq -2 du dr + r^2 d\Omega^2,
\ee
where $r$ is approximately a light-cone coordinate.
The Einstein tensor is then approximated by
\be
G_{ur} \simeq \frac{1}{r^2},
\ee
with all other components vanishing.
There is a singularity at $r = 0$.
But we note that it is a light-like singularity.
}

\section{Geometry of Full Space-Time}

In the above we have discussed the geometry of the region outside the matter shell.
The space enclosed by the collapsing shell is by assumption in vacuum 
(with vanishing energy momentum tensor),
hence it has to be the Minkowski space,
as required by Birkhoff's theorem.

To understand the geometry of the region occupied 
by the collapsing matter,
one can divide the collapsing shell of finite thickness 
into infinitely many infinitesimally thin layers,
separated by infinitesimal gaps \cite{Kawai:2013mda}.
For the inner most layer of dust, 
the space inside is Minkowskian.
Assuming that the dust cannot pass through itself,
the first layer comes in at the speed of light
and piles up at the origin.
The second layer comes in after that,
slowing down at the Schwarzschild radius defined by the first layer.
Then the third layer falls in, and so on.
This description can also be applied to a collapsing solid sphere as well.

One can treat the trajectory of each layer
in a way similar to what was done for $R(u)$ in the above \cite{Kawai:2013mda}.
For a sufficiently large and dense body of collapsing matter,
eventually the radius of each layer 
approaches to its own Schwarzschild radius 
defined for the total mass enclosed by that layer,
until the last stage when the evaporation is
approaching the Planck scale.

The ultimate fate of the collapsing shell at the Planck scale
(whether it evaporates completely)
and the resolution of the singularity at the origin (if any) by a UV-complete theory
are outside the scope of this article.
Semiclassical considerations of the final stage of the evaporation
can be found in e.g. \cite{Kawai:2013mda,FT,O'Loughlin:2013fha}.
The purpose of this article is to resolve puzzles about black holes
at the semiclassical level as much as possible.
It does not exclude
the potential relevance of Planck-scale physics such as string theory to black holes
(e.g. through the idea of the fuzzballs \cite{FuzzBall})
at a more detailed level.

\begin{figure}
\vskip-4em
\includegraphics[scale=0.5]{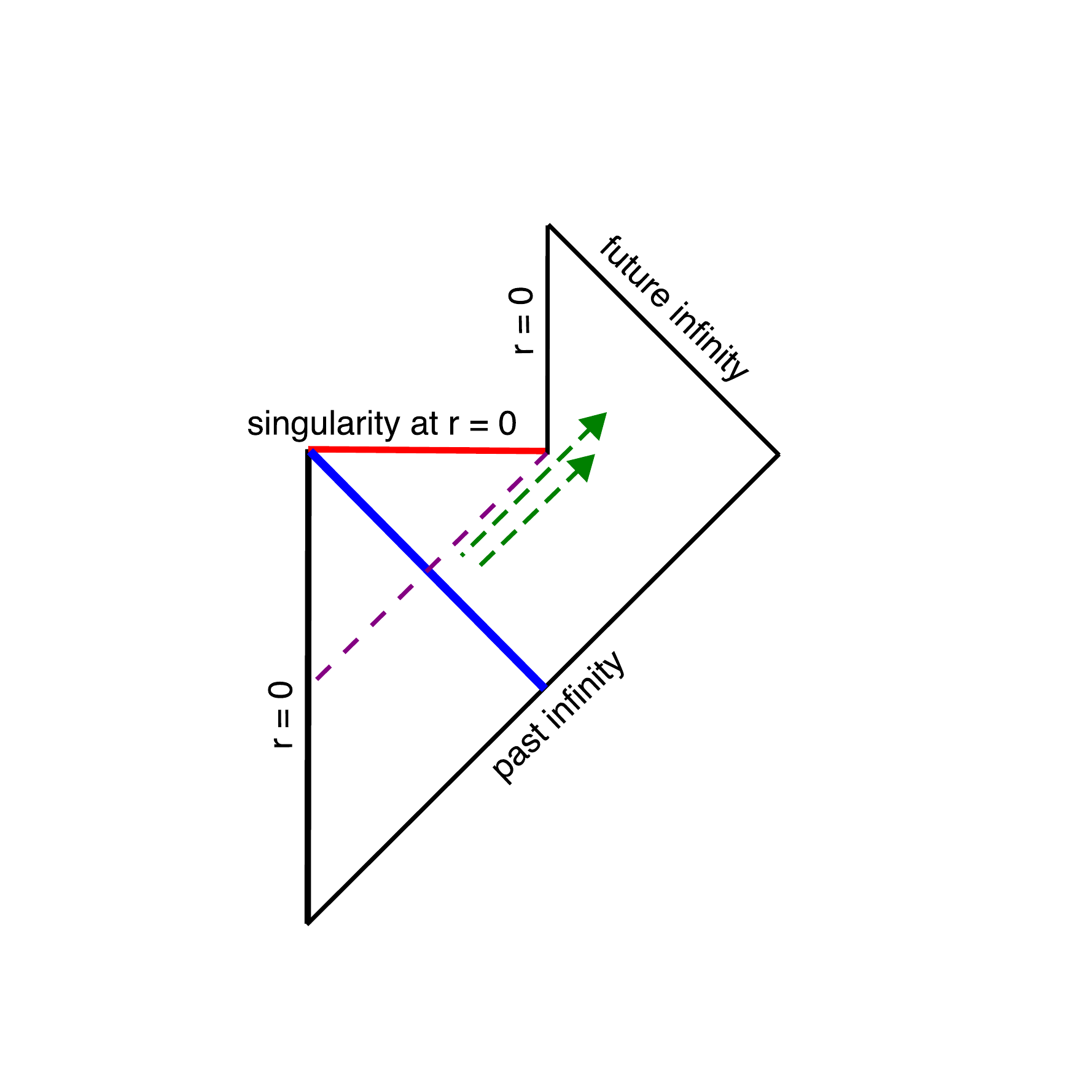}
\includegraphics[scale=0.45]{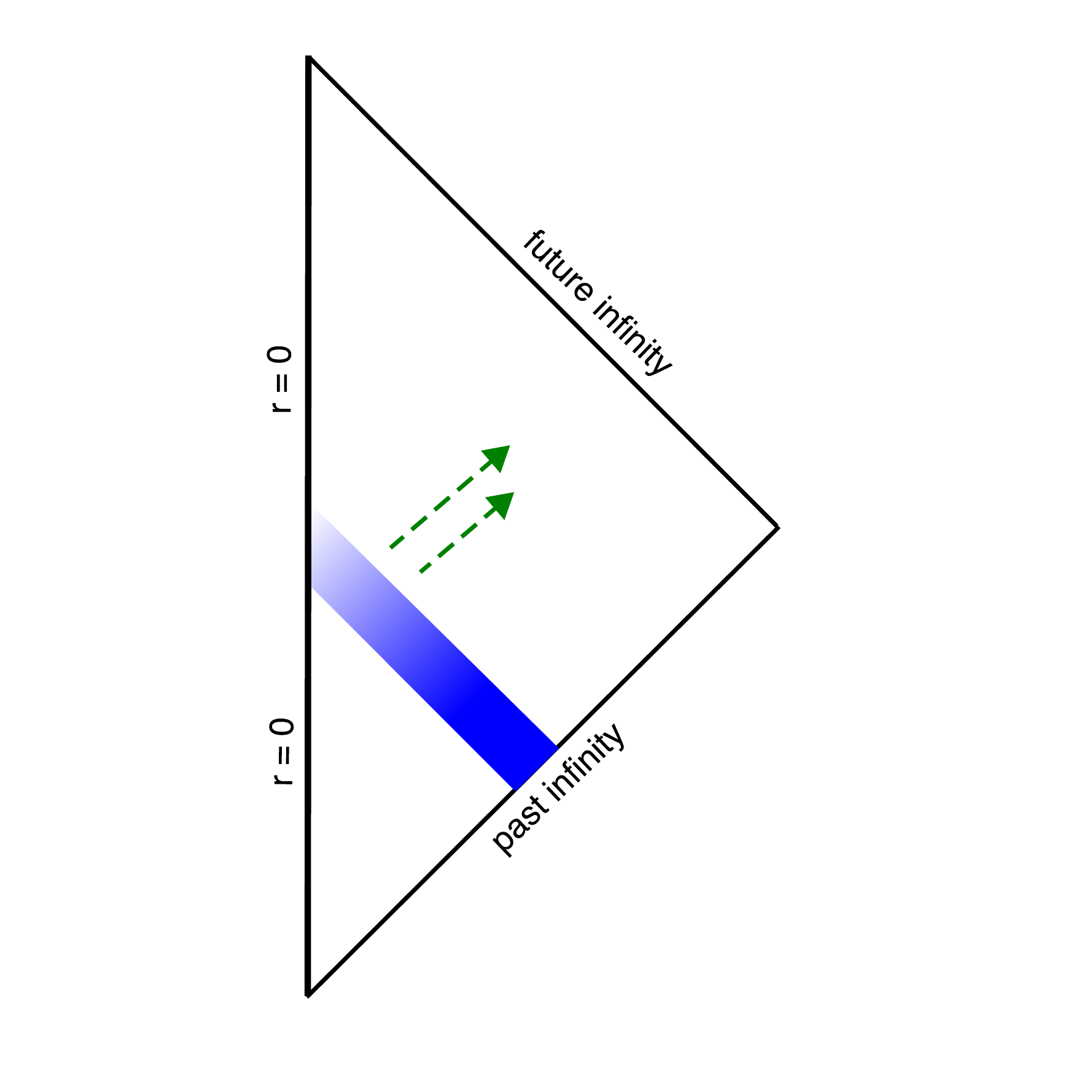}
\vskip-4em
\hskip4cm (a) \hskip9cm (b)
\caption{\small
In both diagrams, 
blue lines represent collapsing null matter shells,
green arrows Hawking radiation.
Figure (a) shows the Penrose diagram for 
the conventional view of a black hole
formed by a collapsing null matter shell.
Figure (b) shows the Penrose diagram for
the collapsing null matter shell 
according to the KMY model.}
\label{Penrose}
\vskip1em
\end{figure}

According to our discussions above,
the whole space-time can be divided into four regions:
\begin{enumerate}
\item
the Minkowskian region inside the shell
with the metric
\be
ds^2 = - dU^2 - 2dU dr + r^2 d\Omega^2.
\ee
\item
the region occupied by the collapsing matter shell.
\item
the region outside the shell 
with the outgoing Vaidya metric (\ref{outgoing-Vaidya}).
\item
the Minkowskian region 
after the shell completely evaporates 
($a(u) = 0$ for $u > u_0$ in the Vaidya metric).
\end{enumerate}
Put together,
they constitute a geodesically complete spacetime.
The corresponding Penrose diagram is depicted in Fig. \ref{Penrose}(b),
along with the Penrose diagram for the conventional view 
in Fig. \ref{Penrose}(a) for comparison.

\hide{
A crucial feature of the Penrose diagram of the KMY model
is that every point in the diagram can be connected to 
the future infinity via outgoing light cones.
In other words,
the complete information of the universe 
can be recorded at the future infinity.
There is no horizon to hide information behind.
}

\section{Infalling Observer}

In principle,
since the distant observer's view is already complete,
there is no need to consider
the viewpoint of an infalling observer.
Nevertheless,
we repeat here the conventional story about how 
an infalling observer passes through the horizon,
and point out the reason why 
it breaks down in the KMY model.

The (static) Schwarzschild solution can be expressed in 
the ingoing Eddington-Finkelstein coordinates as
\be
ds^2 = - \left(1-\frac{a_0}{r}\right)dv^2 + 2dv dr
+ r^2 d\Omega^2,
\ee
where 
\be
v = t + r^*(r).
\ee
The light-like trajectories are given by
\be
\frac{dr}{dv} = - \infty, 
\qquad
\frac{dr}{dv} = 
\frac{1}{2}\left(1-\frac{a_0}{r}\right).
\ee
An infalling trajectory with $\frac{dr}{dv} < 0$
moves at a speed slower than light
as long as $\frac{dr}{dv}$ is finite.

The proper time it takes an infalling observer to reach the Schwarzschild radius
from $r = a_0 + \eps$ $(a_0 \gg \eps > 0)$ is
\be
\int ds = \int_{a_0}^{a_0+\eps} dr \; \sqrt{\left(1-\frac{a_0}{r}\right)\left(\frac{dr}{dv}\right)^{-2} 
- 2\left(\frac{dr}{dv}\right)^{-1}},
\ee
which is finite as long as 
$|\frac{dr}{dv}|^{-1}$ is finite.

Note that $v$ is related to $u$ via
\be
v = u + 2r^*,
\label{v-u}
\ee
where $r^*$ is given in (\ref{r*}),
so that a point on the horizon $r = a_0$ at finite $v$
corresponds to $u = \infty$.
This is why
a distant observer can never see the infalling observer to 
pass the horizon of a static black hole.

It is conventionally assumed that
for a very small Hawking radiation,
there is no reason for the infalling observer
to have a dramatically different experience.
The infalling observer is still expected to 
be able to pass through 
the horizon within finite proper time.

In the KMY model,
the geometry outside the collapsing shell
looks just like that of a black hole.
The effect of a small Hawking radiation also 
makes a small difference to a distant observer.
However, 
the matter shell evaporates away in finite time 
and the space is Minkowkian for $u > u_0$.
A coordinate transformation of the form (\ref{v-u})
implies that 
the infalling observer should fall into
the Minkowski space 
before $r$ reaches the Schwarzschild radius $a(u)$,
because $r \rightarrow a$ implies $r^* \rightarrow - \infty$,
which in turn implies $u \rightarrow \infty$
for a finite $v$,
and we have $a(u) = 0$ for large $u$.

Therefore,
in agreement with the viewpoint of the distant observer,
the infalling observer can never pass through the Schwarzschild radius.
From the viewpoints of both the distant observer and the infalling observer,
the collapsing shell as well as the infalling observer
stay outside the Schwarzschild radius at all times.
They both see the collapsing shell evaporate completely,
and the space-time becomes Minkowskian again
after all of the Hawking radiation dissipates to the infinity.

\section{Information and Firewall}

From the viewpoint of a distant observer,
as a collapsing matter shell gets closer to the Schwarzschild radius,
the shell gets dimmer and looks more like a black hole.
While the matter never falls inside the horizon,
Hawking radiation is created in the neighborhood of the matter.
It is thus natural to assume that 
Hawking radiation carries the information
about the details of the matter shell, 
and the information loss paradox is resolved.

\hide{
There is no horizon to pass through for
an infalling observer to tell a different story.
The black-hole complementarity \cite{Susskind:1993if,'tHooft:1984re} 
is therefore irrelevant.
}

It used to be a common belief that,
for a large black hole,
an infalling observer 
can pass through the horizon without feeling anything 
dramatically different from the ordinary Minkowski space in vacuum.
We find that,
on the contrary,
if the effect of Hawking radiation is consistently taken into account
before the horizon appears,
as it is in the KMY model,
the surface of the collapsing matter stays outside the Schwarzschild radius,
so it is impossible for an infalling observer
to mistakenly assume that 
the space around the Schwarzschild radius is in vacuum.

In fact, 
it is proven as a theorem \cite{Mathur:2009hf} that
the unitarity of quantum mechanics is inconsistent with
the assumption that nothing happens at the horizon,
and fuzzballs \cite{FuzzBall} were proposed to replace the horizon in vacuum.
A more recent proposal for an eventful horizon 
was that of the firewall \cite{Almheiri:2012rt}.
It states that,
in order for the Hawking radiation to be a pure state
(when the collapsing matter comes in a pure state),
there has to be a high energy flux
-- the so-called ``firewall'' --
near the horizon of a sufficiently old black hole.

The firewall is essentially the blue-shifted Hawking radiation 
generated near the horizon,
before it propagates to a distance with a large red shift.
For the KMY model,
we can check how much energy is there 
in the blue-shifted Hawking radiation.
The question is whether 
the presence of the collapsing shell,
which cuts off the near horizon region at a certain distance from $a(u)$,
would reduce the firewall to a cold shower.

At the outer surface of the collapsing shell,
the energy-momentum tensor of the Hawking radiation is
\be
T_{++} = \frac{du}{dx^+}\frac{du}{dx^+} T_{uu}
= - \frac{1}{8\pi}\left(1-\frac{a(u)}{R(u)}\right)^{-1}\frac{\dot{a}(u)}{R^2(u)},
\label{T++}
\ee
where $dx^+$ is the local outgoing light-cone coordinate
normalized by $ds^2 = {dx^+}^2$ along
an outgoing null trajectory.

If the radius $R(u)$ of the shell could get arbitrarily close 
to the Schwarzschild radius $a(u)$,
the energy-momentum tensor $T_{++}$ 
can be arbitrarily large at the collapsing shell 
due to an arbitrarily large blue shift $(1-a/R)^{-1}$.
However,
the formula (\ref{T++}) is valid only for $r \geq R(u)$,
and $R(u)$ always lags behind $a(u)$.
Hence $R(u)$ may never be sufficiently close to $a(u)$
for $T_{++}$ to be incredibly large.
Indeed, 
when the shell radius $R(u)$ is very close to $a(u)$,
it can be approximated by the expression (\ref{R-sol}),
so that
\be
T_{++} \simeq \frac{1}{16\pi} \frac{1}{a^2(u)},
\ee
which is very small for a large mass.
For a shell collapsing at a speed slower than light,
the radius $R(u)$ would be even farther away from $a(u)$,
and $T_{++}$ at $r = R(u)$ would be even smaller.
Our conclusion is thus that 
there is nothing that can be justified to be called a firewall.

\section{Generalization}

The basic ideas involved in the discussions above 
allow straightforward generalizations to higher dimensions.
The outgoing Vaidya metric in $D$-dimensional space-time is
\cite{Iyer:1989nd}
\be
ds^2 = -\left(1 - \frac{2M(u)}{(D-3)r^{D-3}}\right) du^2
- 2 dudr + r^2 d\Omega_{D-2}^2,
\ee
where $d\Omega_{D-2}^2$ is the metric for a $(D-2)$-dimensional unit sphere.
In all dimensions higher than 4,
this metric can be used to describe the region
outside a collapsing null shell
as we did above in 4 dimensions.
The Schwarzschild radius $a(u)=2M(u)/(D-3)$
has a space-like trajectory 
($ds^2 > 0$) as long as the mass $M(u)$ decreases with time,
as it should due to Hawking radiation.

The  only non-vanishing component of the energy-momentum tensor is
\be
T_{uu} = - \frac{(D-2)}{8\pi(D-3)r^{D-2}}\dot{M}(u),
\ee
which can be interpreted as that of the Hawking radiation.

We expect that 
the basic ingredients of the arguments above for no horizon
can be readily applied to black holes in higher dimensions,
and to generalized Vaidya metrics \cite{Wang:1998qx},
as well as other metrics including the effect of Hawking radiation.

\hide{
For a sufficiently large and dense collapsing shell,
the mass distribution at a later stage
has every layer of its mass very close to
the Schwarzschild radius defined by the mass
enclosed in that layer.
Hence the first term in $ds^2$ quadratic in $du$
is nearly zero for every layer,
so that
\be
ds^2 \simeq - 2 dudr + r^2 d\Omega^2,
\ee
and both coordinates $u$ and $r$ are light-like.
The singularity at $r=0$ is thus light-like.
}

\section{Summary}

The traditional view of the formation and evaporation 
of a black hole depicted in Fig. \ref{Penrose-old-1}
is obtained as a composition of two Penrose diagrams:
the Penrose diagram for the formation of a black hole 
without the back reaction of Hawking radiation,
and that for the asymptotically flat spacetime
with Hawking radiation.
The formation of a black hole and its evaporation are 
treated separately as independent processes.

The KMY model, 
on the other hand,
consistently includes the effect of 
the back reaction of Hawking radiation
from the very beginning to the very end.
All infalling time-like or light-like geodesics 
originated from the outside of the collapsing shell
can be extended to be geodesically complete 
without encountering any horizon.
The Penrose diagram for the KMY model in Fig. \ref{Penrose}(b)
shows no horizon and no information loss
at a macroscopic scale.

Furthermore, 
the conventional assumption about an empty horizon is replaced by 
the collapsing matter outside the Schwarzschild radius. 
There is no firewall 
because the blue-shift of the Hawking radiation is cut off
by the radius of the collapsing shell.

For a distant observer,
it takes a longer time for light to reach him or her 
from the collapsing star 
as its outer radius gets closer to the Schwarzschild radius,
hence the star looks darker.
Therefore the fact that there is no horizon 
is not in contradiction with observational evidences of 
phenomenological black holes.

Despite the fact that 
we have ignored non-spherical and massive 
contributions to Hawking radiation,
the qualitative features presented above
should be valid for a generic incipient black hole,
and provide us with a comprehensive semiclassical understanding
of the information loss paradox,
as well as related problems such as the firewall.

\section*{Acknowledgement}

The author would like to thank 
Hikaru Kawai for sharing his original ideas
on which this work is based.
He also thanks 
Heng-Yu Chen, Kazuo Hosomichi, Takeo Inami, Keisuke Izumi, 
Samir Mathur and Chih-Hung Wu for discussions.
The work is supported in part by
the Ministry of Science and Technology, R.O.C.
and by National Taiwan University.


\end{CJK} 
\end{document}